\documentclass[12pt]{article}
\usepackage{amsmath,amssymb}
\newtheorem{thm}{Theorem}[section]

\newtheorem{assumption}[thm]{Assumption}

\newtheorem{definition}[thm]{Definition}

\newtheorem{example}[thm]{Example}

\newtheorem{remark}[thm]{Remark}

\newtheorem{protocol}[thm]{Protocol}

\renewcommand{\O}{\mathcal{O}}
\newcommand{\Z}{\mathbb{Z}}

\newcommand{\md}[1]{\,\, {\rm (mod\,\, #1) }\,}

\newcommand{\Section}[1]{\section{#1}\setcounter{equation}{0}}

\title{Remarks on MOBS and cryptosystems using semidirect products
}

\date{Draft of \today}%
\author{Chris Monico \\
  {\small Department of Mathematics and Statistics\vspace{-2mm}}\\
  {\small Texas Tech University\vspace{-2mm}}\\
  {\small {\em e-mail:\/} c.monico@ttu.edu }
  }

\begin{document}\maketitle
\thispagestyle{empty}
\begin{abstract}
Recently, several cryptosystems have been proposed based
semidirect products of various algebraic structures \cite{habe13, kahr16, grig19, rahm21}.
Efficient attacks against several of them have already
been given \cite{kreu14, monico20, rudy21, brown21, batt21},
along with a very general attack in \cite{romankov16}.
The purpose of this note is to provide an observation that can be used 
as a point-of-attack
for similar systems, and show how it can be used to efficiently
cryptanalyze the MOBS system.
\end{abstract}

\Section{General semidirect product cryptosystems}        \label{sec:intro}

  In this section, we describe the general framework
encompassing several recently proposed algebraic cryptosystems,
including \cite{habe13, kahr16, grig19, rahm21}, and give 
a general observation which applies to them all. 
That observation will be used in the next section to give
a polynomial-time attack on the proposed MOBS system \cite{rahm21}.

Suppose that $G$ is a semigroup and $S$ is a sub-semigroup of
endomorphisms of $G$. One can define the semidirect product
$G\rtimes S$ as the set $G\times S$ together with the operation
\[
  (g_1, \phi_1)(g_2, \phi_2) = (\phi_2(g_1)g_2,\,\, \phi_1\circ\phi_2).
\]
One can then build a Diffie-Hellman-like key exchange protocol
as follows.

\begin{enumerate}
  \item[(i)] Alice and Bob agree on an element $(g,\phi)\in G\rtimes S$.
  \item[(ii)] Alice chooses a private integer $a$, computes 
             $(g,\phi)^a = (A, \phi^a)$, and sends $A$ to Bob.
  \item[(iii)] Bob chooses a private integer $b$, computes
             $(g,\phi)^b = (B, \phi^b)$, and sends $B$ to Alice.
  \item[(iv)] Alice computes $K_A = \phi^a(B)A$.
  \item[(v)] Bob computes $K_B = \phi^b(A)B$.
\end{enumerate}
Since 
\[
  (K_A, \phi^{a+b}) = (B, \phi^b)(A, \phi^a) = (g,\phi)^{b+a} = (A,\phi^a)(B,\phi^b) = (K_B, \phi^{a+b}),
\]
it follows that $K_A=K_B$, so this is Alice and Bob's shared secret key, $K$.
One also has that
\begin{eqnarray*}
  A &=& \phi^{a-1}(g)\phi^{a-2}(g)\cdots \phi(g)g,\\
  B &=& \phi^{b-1}(g)\phi^{b-2}(g)\cdots \phi(g)g, \hspace{12pt}\mbox{ and }\\
  K &=& \phi^{a+b-1}(g)\phi^{a+b-2}(g)\cdots \phi(g)g.
\end{eqnarray*}

In general, it is not necessary for an attacker Eve to determine $a$ or $b$ to 
recover the shared key $K$. It would be sufficient for her to find an
endomorphism $\psi$ of $G$ which commutes with $\phi$ and satisfies
\begin{equation} \label{eqn:ss}
  \psi(g)A = \phi(A)g.
\end{equation}
If she can find such an endomorphism, it follows that
\begin{eqnarray*}
  \psi(B)A = \psi\left( \prod_{j={b-1}}^0 \phi^j(g) \right) A 
           = \left( \prod_{j={b-1}}^0 \phi^j(\psi(g))\right) A 
           &=& \left( \prod_{j={b-1}}^1 \phi^j(\psi(g))\right) \psi(g) A \\
           &=& \left( \prod_{j={b-1}}^1 \phi^j(\psi(g))\right) \phi(A) g \\
           &=& \left( \prod_{j={b-1}}^2 \phi^j(\psi(g))\right) \phi(\psi(g)A) g \\
           &=& \left( \prod_{j={b-1}}^2 \phi^j(\psi(g))\right) \phi^2(A) \phi(g) g \\
           &\vdots & \\
           &=& \phi^b(A)B = K.
\end{eqnarray*}

\Section{MOBS} \label{sec:MOBS}
  In \cite{rahm21}, the authors propose the following. Let $k$
  be a positive integer and let $\mathcal{B}_k$ denote the semiring
  of bitstrings of length $k$ (i.e., $\mathcal{B}_k = \Z_2^k$, as a set),
  together with the operations of bitwise OR and bitwise AND.
  It's easy to see that AND distributes over OR and both operations are
  associative, so $\mathcal{B}_k$ with these operations is indeed
  a semiring. Then $G$ will be the multiplicative 
  semigroup of $n\times n$ matrices over $\mathcal{B}_k$.
  
  A permutation $\sigma\in S_k$ naturally acts on $\mathcal{B}_k$
  by permuting the bits, and this extends to an action on $G$.
  The semigroup of endomorphisms $S$ is taken as the symmetric
  group $S_k$; in fact, this is a group of automorphisms of $G$.

  Suppose that $g,\phi, A,$ and $B$ are as in the previous
  section with this choice of $G$ and $S$.
  We will now show how to produce an endomorphism $\psi$
  which commutes with $\phi$ and satisfies \eqref{eqn:ss}.
  In fact, we will determine an integer $\alpha$ for which
  \[
    \phi^\alpha(g)A = \phi(A)g.
  \]
  First note that such an $\alpha$ necessarily exists, since
  Alice's integer $a$ satisfies this.

  Since $\phi$ is a permutation on $\{1,2,\ldots, k\}$,
  we can determine its disjoint cycle decomposition
  $\phi = \sigma_1\cdots\sigma_t$ with $\O(k)$ operations.
  Since the cycles $\sigma_1,\ldots, \sigma_t$ are disjoint,
  they commute, and so $\phi^\alpha(g)A = \phi(A)g$
  if and only if
  \[
    \left( \sigma_1^\alpha \cdots \sigma_t^\alpha\right)(g)A = \phi(A)g.
  \]
  For each $j$, one can find an integer $\alpha_j$
  for which $\sigma_j^{\alpha_j}(g)A$ agrees with $\phi(A)g$
  in the bit positions corresponding to that cycle (i.e., the
  orbit of $\sigma_j$ which has length greater than 1).
  This can be done with brute force by computing
  $gA, \sigma_j(g)A, \sigma_j^2(g)A,\ldots$ until such an
  $\alpha_j$ is found. This requires that we compute
  at most $|\sigma_j|$ permutation products and matrix products.

  Then use the Chinese Remainder Theorem to find an
  integer $\alpha$ for which $\alpha\equiv \alpha_j \md{ |\sigma_j|}$
  for all $j$. It follows that $\phi^\alpha(g)A = \phi(A)g$.

  Since $|\sigma_1|+\cdots + |\sigma_t| \le k$, we have to 
  compute no more than $k$ permutation products and matrix products.
  Since these operations are polynomial-time in the key size, and
  $k$ is less than the key size, it follows that this is polynomial-time.
  The final Chinese Remainder Theorem calculation solves a system 
  of congruences with moduli $|\sigma_1|,\ldots, |\sigma_t|$.
  If $N=\prod |\sigma_j|$, then the size of $N$ is about
  $\log{N} = \sum \log|\sigma_j| \le k$, and this can be
  done using $\O(\log^2{N}) = \O(k^2)$ operations \cite{bach97},
  so it is also polynomial-time.
  
  We extended the Python code generously made available by the
  authors of \cite{rahm21} to implement this attack, and ran
  experiments for various values of $k$ (of the same form they suggested,
  being a sum of the first several primes). For each indicated value
  of $k$, we used $n=3$ ($3\times 3$ matrices) and generated 20
  shared keys. We report the average wall-clock time to recover
  each shared key on a single core of an i7 processor at 3.10GHz.

  \begin{center}\begin{tabular}{r|r}
    $k$ & Avg. time (seconds) \\
    \hline
    100 & 0.0878 \\
    197 & 0.2374 \\
    381 & 0.5325 \\
    791 & 1.7000
  \end{tabular}\end{center}

\bibliographystyle{plain}

\end{document}